\documentclass[twocolumn,english]{aastex631} 

\RequirePackage[normalem]{ulem} 
\RequirePackage{color}\definecolor{RED}{rgb}{1,0,0}\definecolor{BLUE}{rgb}{0,0,1} 
\providecommand{\DIFaddbegin}{} 

\begin{document}
\title{Pre-flight Background Estimates for COSI}

\author[0000-0002-2664-8804]{Savitri Gallego}
\affiliation{Institut für Physik \& Exzellenzcluster PRISMA+,Johannes Gutenberg-Universität Mainz, 55099 Mainz, Germany. email : sgallego@uni-mainz.de}

\author[0000-0001-8160-5498]{Uwe Oberlack}
\affiliation{Institut für Physik \& Exzellenzcluster PRISMA+,Johannes Gutenberg-Universität Mainz, 55099 Mainz, Germany. email : sgallego@uni-mainz.de}

\author[0009-0004-9049-2199]{Jan Lommler}
\affiliation{Institut für Physik \& Exzellenzcluster PRISMA+,Johannes Gutenberg-Universität Mainz, 55099 Mainz, Germany. email : sgallego@uni-mainz.de}

\author[0000-0002-6774-3111]{Christopher M. Karwin}
\affiliation{Department of Physics and Astronomy, Clemson University, Clemson, SC 29634, USA}

\author[0000-0002-0313-8852]{Francesco Fenu}
\affiliation{Agenzia Spaziale Italiana, Via del Politecnico s.n.c., 00133 Roma, Italy}

\author[0000-0002-6082-5384]{Valentina Fioretti}
\affiliation{INAF, Osservatorio di Astrofisica e Scienza dello Spazio (OAS) di Bologna, Via P. Gobetti 93/3, 40129 Bologna, Italy}

\author[0000-0001-9067-3150]{Andreas Zoglauer}
\affiliation{Space Sciences Laboratory, UC Berkeley, 7 Gauss Way, University of California, Berkeley, CA 94720, USA}

\author[0000-0003-2317-9560]{F. M. Follega}
\affiliation{University of Trento, Department of Physics, V. Sommarive 14, I-38123, Trento, Italy}
\affiliation{INFN-TIFPA, V. Sommarive 14, I-38123 Trento, Italy}

\author[0000-0001-5603-3950]{A. Perinelli}
\affiliation{University of Trento, Department of Physics, V. Sommarive 14, I-38123, Trento, Italy}
\affiliation{INFN-TIFPA, V. Sommarive 14, I-38123 Trento, Italy}

\author[0000-0002-5808-7239]{Roberto Battiston}
\affiliation{University of Trento, Department of Physics, V. Sommarive 14, I-38123, Trento, Italy}
\affiliation{INFN-TIFPA, V. Sommarive 14, I-38123 Trento, Italy}

\author[0000-0001-5038-2762]{Roberto Iuppa}
\affiliation{University of Trento, Department of Physics, V. Sommarive 14, I-38123, Trento, Italy}
\affiliation{INFN-TIFPA, V. Sommarive 14, I-38123 Trento, Italy}

\author[0000-0001-9567-4224]{Steven E. Boggs}
\affiliation{Department of Astronomy \& Astrophysics, UC San Diego, 9500 Gilman Drive, La Jolla, CA 92093, USA}

\author[0009-0005-1439-7808]{Saurabh Mittal}
\affiliation{University Würzburg, D-97074 Würzburg, Germany}

\author[0000-0002-1757-9560]{Pierre Jean}
\affiliation{Institut de Recherche en Astrophysique et Planétologie, Université de Toulouse, CNRS, CNES, UPS, 31028 Toulouse, Cedex 4, France.}

\author[0000-0001-6677-914X]{Carolyn Kierans}
\affiliation{NASA Goddard Space Flight Center, Greenbelt, MD 20771, USA}

\author[0000-0002-8028-0991]{Dieter H. Hartmann}
\affiliation{Department of Physics and Astronomy, Clemson University, Clemson, SC 29634, USA}

\author[0000-0001-5506-9855]{John A. Tomsick}
\affiliation{Space Sciences Laboratory, UC Berkeley, 7 Gauss Way, University of California, Berkeley, CA 94720, USA}

\collaboration{20}{(on behalf of the COSI Collaboration)}

\begin{abstract}
    The Compton Spectrometer and Imager (COSI) is a Compton telescope designed to survey the 0.2 - 5 MeV sky, consisting of a compact array of cross-strip germanium detectors. It is planned to be launched in 2027 into an equatorial low-Earth (530 km) orbit with a prime mission duration of 2 years. The observation of MeV gamma rays is dominated by background.
    Thus, background simulation and identification are crucial for predicting the sensitivities of instruments.
In this work we perform Monte Carlo simulations of the background for the first 3 months in orbit, and we extrapolate the results to 2 years in orbit, in order to determine the build-up of the  activation due to long-lived isotopes.
The simulations account for the known background components, and include time-dependent \DIFaddbegin rate variations due to the geomagnetic cut-off, South Atlantic Anomaly (SAA) passages. In addition, they include detailed modeling of the delayed activation due to short and long-lived isotopes. 
We determine the rates of events induced by the background that are reconstructed as Compton events in the simulated COSI data. We find that the extragalactic background photons dominate at low energies ($<$660 keV), while delayed activation from cosmic-ray primaries (proton/alpha) and albedo photons dominate at higher energies. As part of this work, a comparison at low latitude ($|b|\leq$1°) between recent measurement of the SAA 
and the AP9/AE9 model has been made, showing an overestimation of the flux by a factor $\sim$9 by the model. The systematic uncertainties associated with these components are quantified.

\end{abstract}


\section{Introduction} \label{sec:intro}

The soft to medium gamma-ray regime ($\sim 0.1-30$~MeV) is one of the least explored energy ranges across the electromagnetic spectrum, with instrumental sensitivities in this energy band that is orders of magnitude worse than neighboring energy bands. One of the main reasons for this is that detectors that operate in this  band suffer from high instrumental backgrounds \citep{COMPTELbck,RHESSI,SPIlines2,SPIlines} . The interactions of cosmic ray and albedo particles not only induce a prompt instrumental background when they deposit energy in the detectors, but they can also induce a delayed background due to the activation of the instrument materials. Thus, background simulation and identification are crucial for predicting the sensitivities of instruments.

In this work we present the pre-flight background estimates for the Compton Spectrometer and Imager (COSI), a Small Explorer (SMEX) satellite mission scheduled to launch  in 2027 into a low-Earth low-inclination orbit~\citep{tomsick2023comptonspectrometerimager}. COSI is a Compton telescope which operates as a wide-field imager, spectrometer, and polarimeter. One of COSI's strengths is its good energy resolution, having a full-width half-maximum (FWHM) of 6.0~keV (9.0~keV) at 0.511~MeV (1.157~MeV). Sixteen high-purity cross-strip germanium semiconductor detectors (each 8 × 8 × 1.5 cm$^3$) are arranged in a 2 × 2 × 4 array that is sensitive to photons between 0.2 to 5 MeV. The detectors are surrounded on four sides and the bottom by active shields. The scintillator material used is bismuth germanium oxide (BGO), and they are read out by silicon photomultipliers (SiPMs). The shields have several functions, including passive background attenuation, active anticoincidence for background radiation coming from the sides and bottom, anticoincidence of gamma-rays that escape from the germanium detectors, and finally they serve as a gamma-ray transient monitor.

The paper is structured as follows. In Section~\ref{sec:bcksimu}, we present the input models and software used for the background simulations. The simulation results are presented in Section~\ref{sec:result} and discussed in Section~\ref{sec:Discussion}. Finally, we give our summary and conclusions in Section~\ref{sec:conclusion}.

\section{Background simulation}\label{sec:bcksimu}

\subsection{Cosmic-ray and albedo components}
\begin{figure}[!ht]
    \centering
    \includegraphics[width=\linewidth]{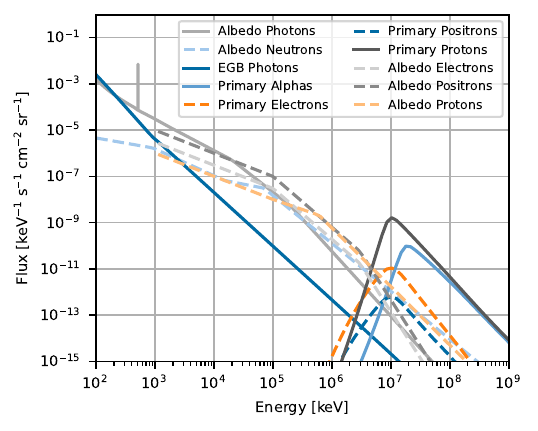}
    \caption{Full on-orbit BG spectra for charged particles, neutrons, and photons. All spectral
components are divided by the solid angle of their region of origin. The average rigidity cutoff (12.6 ~GV) for COSI’s orbit has been used to generate the spectra (see section~\ref{sec:GC}).}
    \label{fig:DC3input}
\end{figure}

The input spectra of cosmic rays and albedo components are generated using the tool from~\cite{Cumani_2019}. We employ an updated version of the code\footnote{\url{https://github.com/cositools/cosi-background/tree/DC3}}, which includes minor code corrections and more recent data. The user can choose the main parameters, including altitude, inclination, geomagnetic cut-off and solar activity. Solar modulation for cosmic-ray protons and alpha particles is accounted for by using the \href{https://www.helmod.org/index.php}{HelMod} online tool \citep{BOSCHINI2018} under the assumption of a data acquisition in the period March – June 2027. The extrapolation to the future solar activity and corresponding solar modulation is done following \cite{Boschini2023}. Cosmic-ray electrons and positron fluxes have been extrapolated from the values measured with AMS-02 ~\citep{AMS2021} to the values expected in 2027 since no modulation is calculated in HelMod for such primaries. The fluxes are calculated with the following formula:
\begin{equation}
\phi_{e^{\pm}}^{27} = \phi_{e^{\pm}}^{AMS21} \frac{\phi_{p}^{27}}{\phi_{p}^{AMS21}} 
\label{eq:AMS}
\end{equation}
where $\phi_{e^{\pm}}^{27}$ is the extrapolated flux of undifferentiated electrons and positrons in 2027, $\phi_{e^{\pm}}^{AMS21}$ is the electron-positron flux as published by the AMS collaboration in 2021, $\phi_{p}^{27}$ is the extrapolated proton flux in 2027 as discussed above and $\phi_{p}^{AMS21}$ is the AMS proton flux. The ratio from equation~\ref{eq:AMS} can be used as a proxy for the modulation evolution since proton fluxes change in time in a similar manner as electrons and positrons (see~\cite{TempElectron,TempPositron} for more details). The albedo photon and neutron spectra and fluxes, as well as the extragalactic gamma-ray background (EGB) component are taken from~\cite{Cumani_2019}. In addition to the albedo photons, the 511~keV line contribution from the atmosphere is taken from~\cite{Harris511lineatm} and rescaled to COSI's altitude.
The resulting spectra are shown in Figure~\ref{fig:DC3input}. All of the input files used for the simulations, will be uploaded in the COSI-tools github repository\footnote{\url{https://github.com/cositools/cosi-sim/tree/main/cosi_sim/Source_Library/DC4/backgrounds}} during spring 2026. Simulations are using the energy range 10$^2$-10$^9$~keV for these components.

\subsection{Galactic diffuse emission}

\begin{figure}
    \centering
    \includegraphics[width=\linewidth]{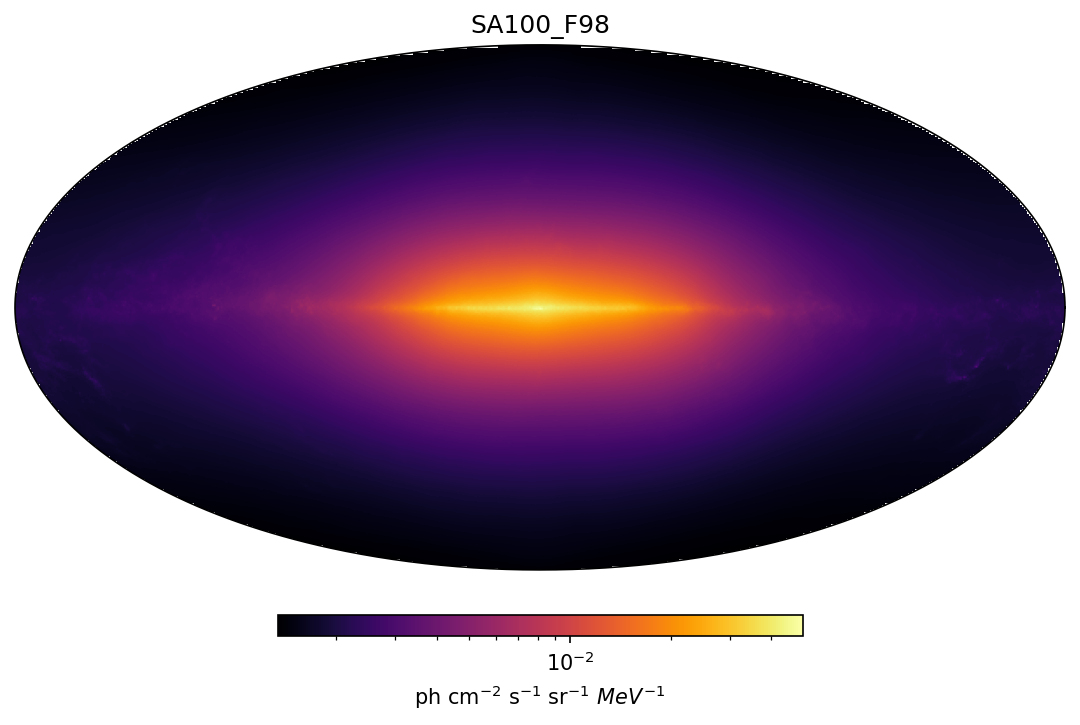}
    \caption{All-sky map of the SA100-F98 model for the Galactic diffuse emission background component. }
    \label{fig:GALPROP}
\end{figure}

\begin{figure*}[!ht]
    \centering
    \includegraphics[trim=100 100 90 100,clip,width=0.30\linewidth]{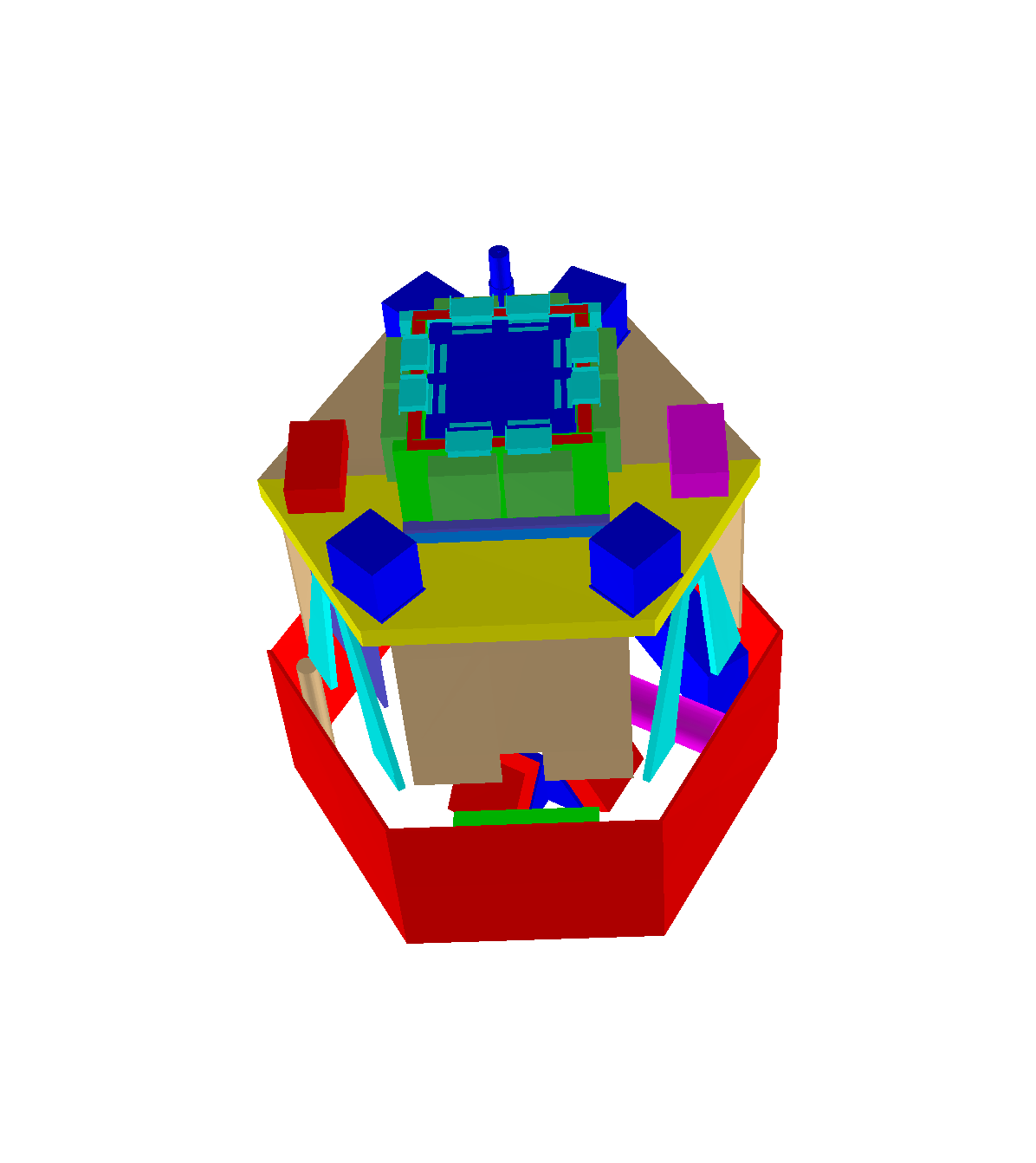}
    \includegraphics[width=0.30\linewidth]{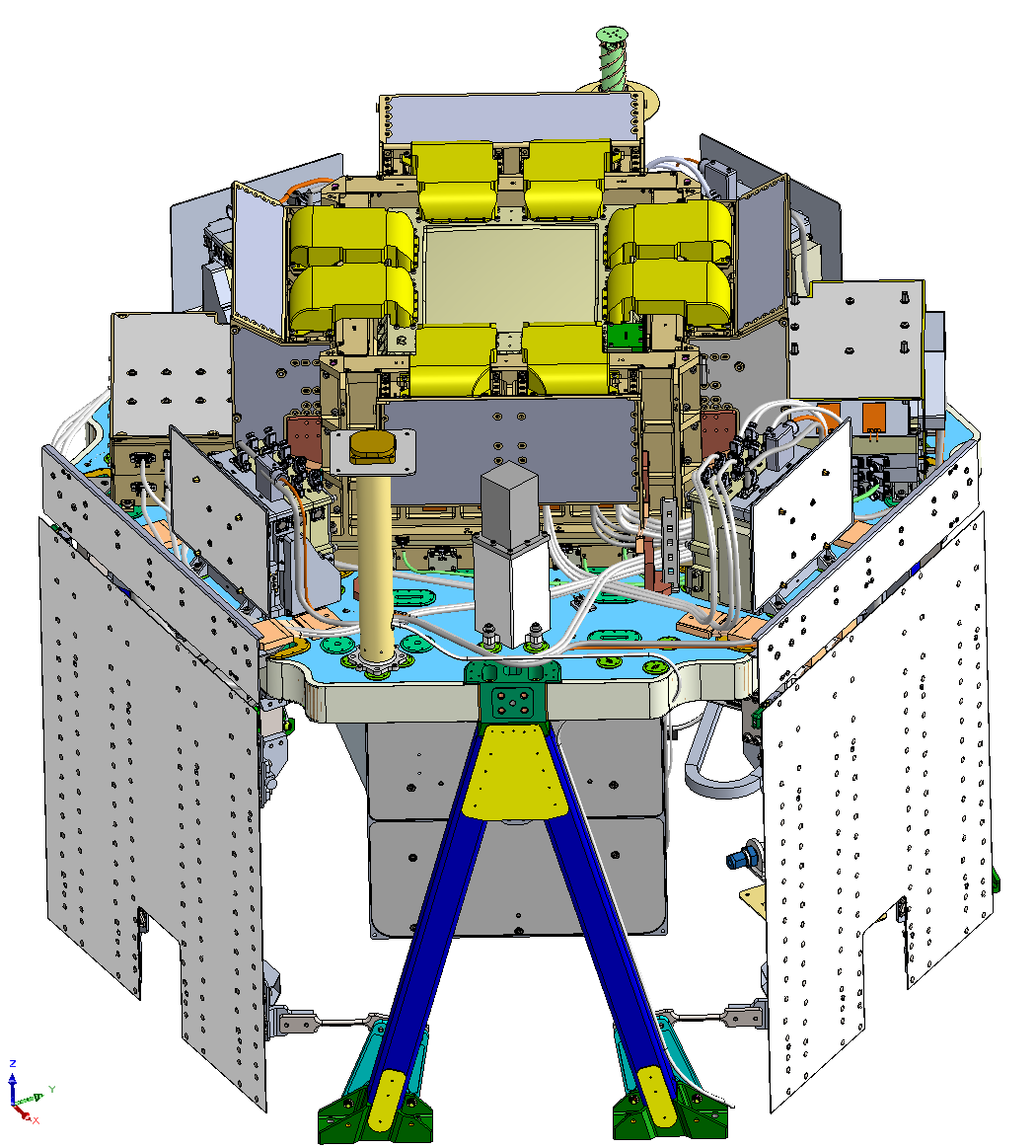}

    \caption{\textbf{Left:} Image of the MEGAlib's mass-model used for the simulation. \textbf{Right:} CAD drawing (version 2025) of COSI payload for comparison. }
    \label{fig:massmodel}
\end{figure*}

The Galactic diffuse continuum emission is modeled using the v57 release of the GALPROP cosmic-ray (CR) propagation and interstellar emissions framework~\citep{Porter_2022}. 
There are six models in total, categorized according to the CR source and interstellar radiation field (ISRF) model used for the prediction. There are 3 CR source models (SA0, SA50, SA100) and two ISRF models (R12, F98). 
These GALPROP models include the total emission, which is dominated by inverse Compton radiation, but also has a small contribution from Bremsstrahlung towards the upper energy bound. For this work we simulate the SA100-F98 model in the energy range 10$^2$-10$^4$~keV as the representative case. The all-sky intensity map (at 501 keV) is shown in Figure~\ref{fig:GALPROP}. The variations coming from the 6 different models are pretty minimal (see Figure 3 in~\cite{Karwin_2023}).

\subsection{MEGAlib software}\label{sec:MEGAlib}

\begin{figure*}[!ht]
    \centering
    \includegraphics[width=0.48\linewidth]{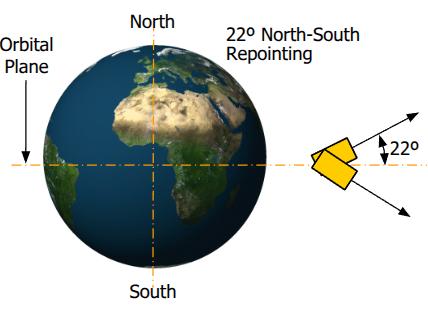}
    \includegraphics[width=0.48\linewidth]{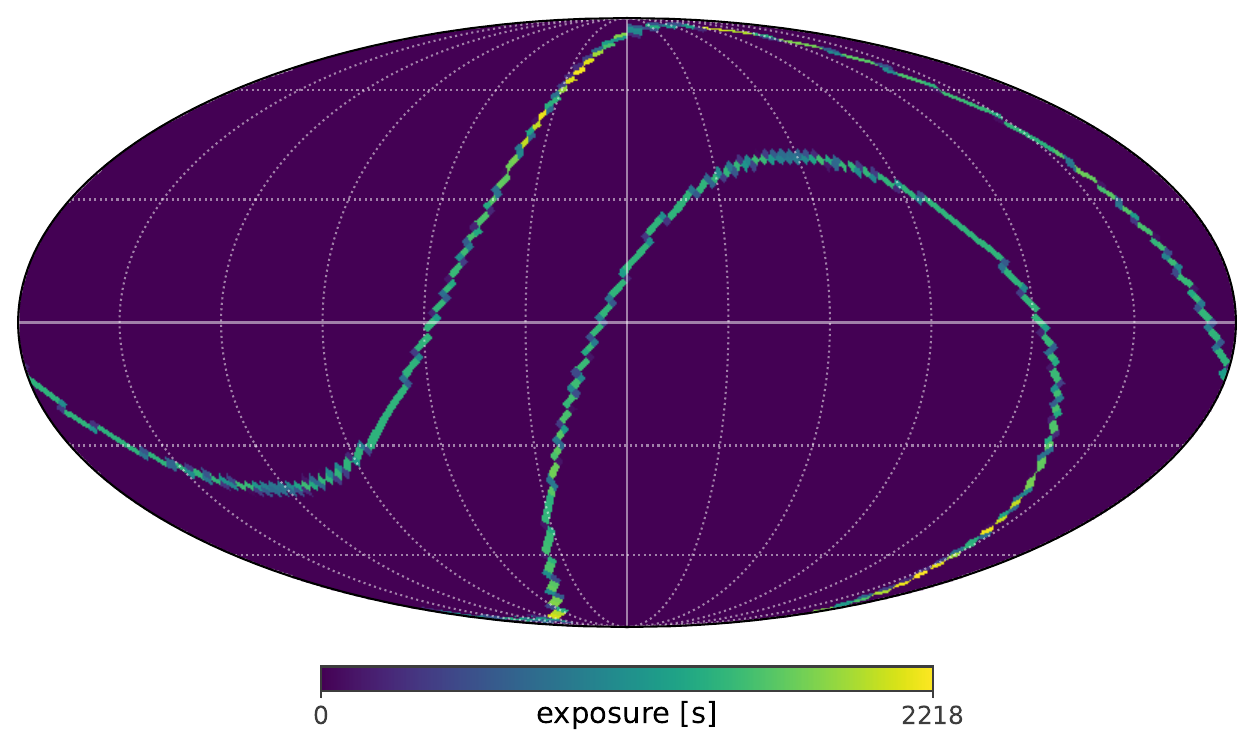}

    \caption{\textbf{Left:} Schematic showing the 22° rocking occurring every 12~hours. \textbf{right:} Pointing of the instrument's z-axis as a function of Galactic latitude and longitude for the 3 months of orbit.}
    \label{fig:cosiorbit}
\end{figure*}

The simulations employ the Medium-Energy Gamma-ray Astronomy library (MEGAlib) software package (version 4.91)~\footnote{\url{https://github.com/zoglauer/megalib/tree/develop-cosi}}, a standard tool in MeV astronomy~\citep{ZOGLAUER2006629}. \textit{Cosima} is the simulation part of MEGAlib, based on GEANT4 v11.2~\citep{GEANT4}. It is able to simulate the emission from a source, the passage of particles through the spacecraft, and their energy deposits in the detector. MEGAlib also performs event
reconstruction, imaging analysis and general high-level analysis (i.e., spectra, light curves, etc.). The mass model used for the simulation is shown in Figure~\ref{fig:massmodel}.

The MEGAlib simulation includes the activation of the instrument induced by CR interactions. Currently, two options are available for this. The first method is a three step process. First, the initial particles and photons are simulated, which determines the prompt interactions and a list of the created isotopes are stored. The activity of each isotope is then calculated based on a given time of irradiation in orbit. The final step simulates the isotope decay at random positions within the mass model volumes in which they were created. The second method of activation simulations keeps in memory each isotope created during the simulation until it decays, with the expected decay time given by the lifetime. This latter method accurately simulates the build-up of the activation during the flight, and it has been used for this study. Given the substantial computational resources needed for such simulations, only the first 3 months of orbit are simulated.

For this work, a basic detector effects engine (DEE) in MEGAlib is applied to the simulated data to account for detector and readout effects. These effects include vetoing events that interact in the shields, charge transport, binning of energy deposits into strips, energy resolutions and thresholds, depth resolution as well as trigger and veto conditions.

The events are then reconstructed using MEGAlib's \textit{revan}~\citep{revan,ZOGLAUER2006629}. A minimum distance of 0.25~cm between any interactions is required for the Compton events. These background simulations were performed as part of the fourth Data Challenge~\footnote{\url{https://doi.org/10.5281/zenodo.15126188}} (DC4) that will be released by the COSI collaboration in spring 2026. The purpose of these data challenges is to aid in the pipeline development, and to familiarize the astrophysics community on how to analyze COSI data.

\subsection{Geomagnetic Cut-Off Dependencies}\label{sec:GC}

The characteristics of the background components composed of charged particles are modified by the geomagnetic field, which varies along the orbit. The spacecraft coordinates have been generated with SPENVIS\footnote{\url{https://www.spenvis.oma.be/intro.php}}, for 3 months of orbit, an altitude of 530~km, and an inclination of $0^\circ$\footnote{COSI will be launched by SpaceX with an expected inclination of 0.25° $\pm$ 0.15°}. In addition, we added a $\pm$22° rocking angle, with the northern sky observed for 12 hours, followed by a 8 minute slewing time, and then the southern sky observed for 12 hours (see Figure~\ref{fig:cosiorbit}). The time dependence of the instrument’s pointing on the sky is simulated with MEGAlib using this orientation.

\begin{figure}
    \centering
    \includegraphics[width=\linewidth]{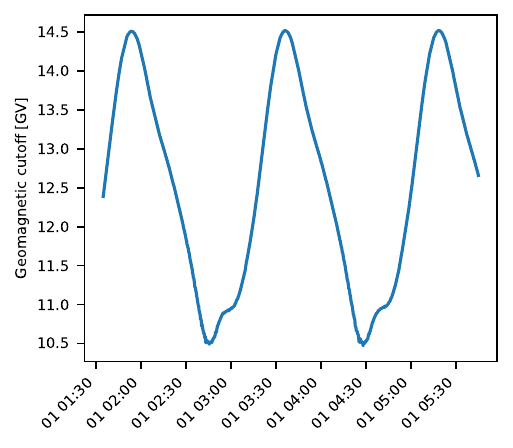}
    \caption{$R_{\mathrm{cutoff}}$ distributions as a function of time for a few hours of COSI orbit.}
    \label{fig:Cutoff}
\end{figure}

\begin{figure}[!h]
    \centering
    \includegraphics[trim=0 0 25 18,clip,width=\linewidth]{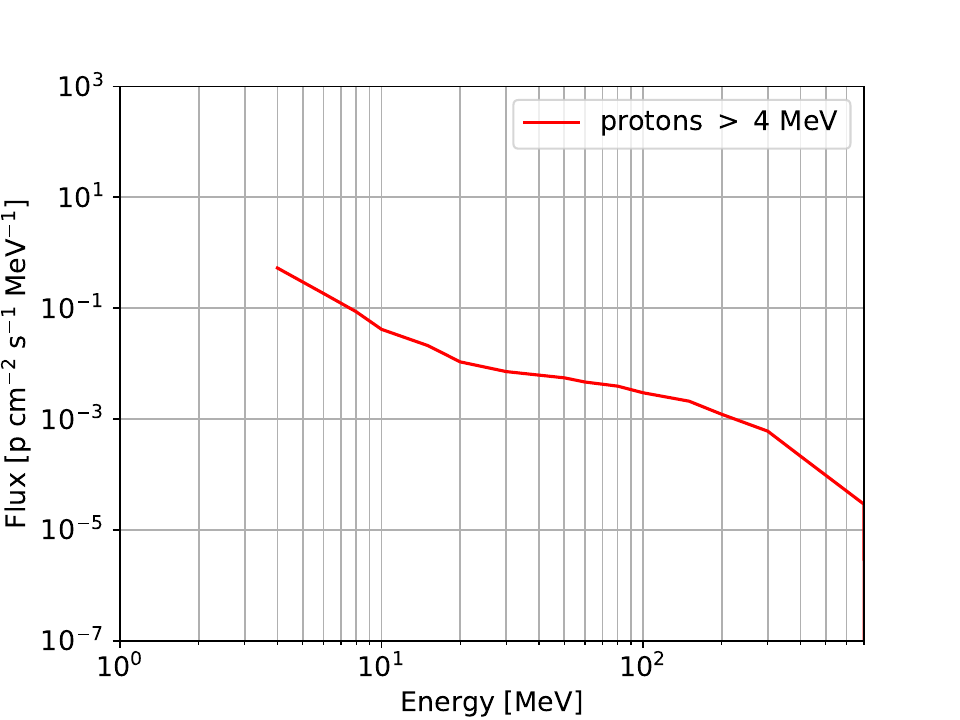}
    \caption{AP9 trapped proton spectrum during SAA passage.}
    \label{fig:SAAinputs}
\end{figure}

\begin{figure}[!h]
     \includegraphics[width=\linewidth]{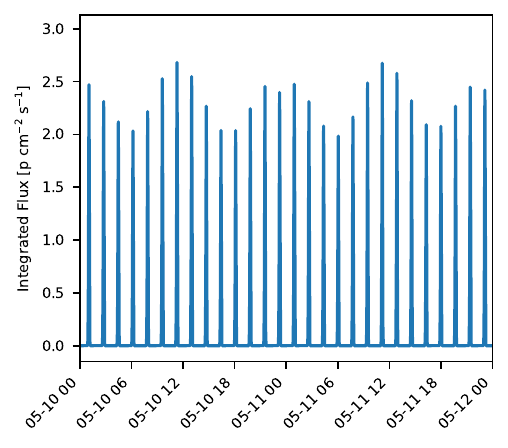}
    \caption{Time profile of the AP9 integrated trapped protons flux between 4 and 2000~MeV for a few days of orbit (2027/05/10 to 2027/05/12) . The variation comes from how COSI crosses the SAA as a function of time, according to AP9.}
    \label{fig:SAALC}
 \end{figure}

\begin{figure}[!h]
    \centering
    \includegraphics[trim=0 0 25 20,clip,width=\linewidth]{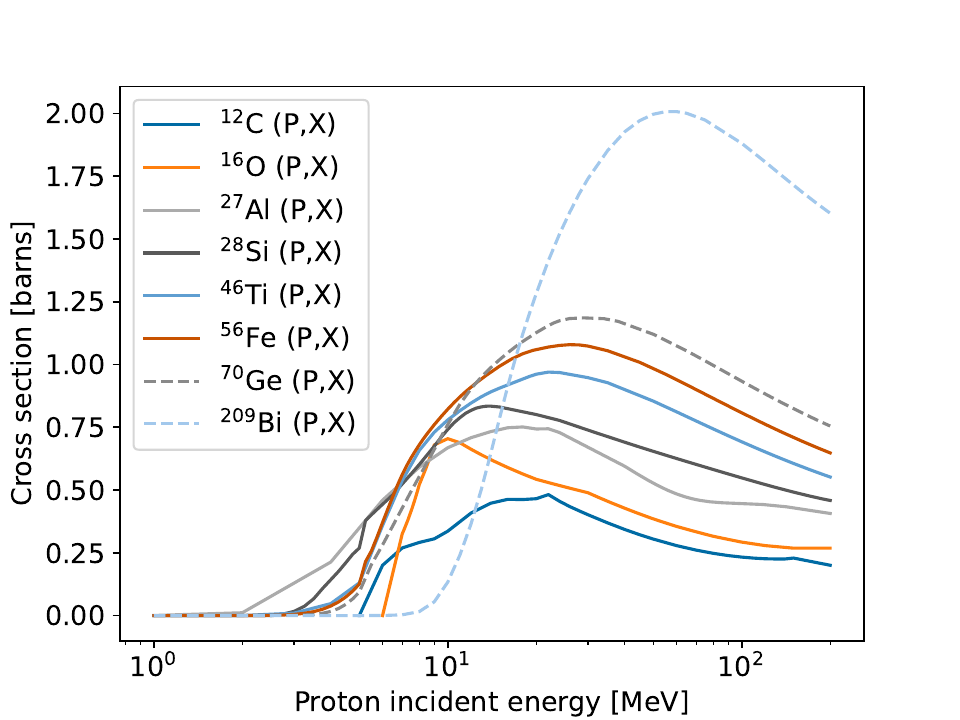}
    \caption{Proton cross sections for different target elements~\citep{JENDL5}.}
    \label{fig:protoncross}
\end{figure}

Based on these coordinates, the geomagnetic cutoff ($R_{\mathrm{cutoff}}$) (in GV) is computed using the python tool OTSO~\citep{OTSO}. We calculate the $R_{\mathrm{cutoff}}$ for each time interval (15~s) of the 3 month orientation file, and use this to obtain the integrated spectrum for each background component as function of time. The $R_{\mathrm{cutoff}}$ dependencies for each component are described in~\cite{Cumani_2019}. The time dependence of the particle fluxes due to $R_{\mathrm{cutoff}}$ variation is handled by including a light curve in the simulations. Note that \textit{cosima} only takes the light curve amplitude into account, and the overall flux normalization is set by the spectrum. We make the simplifying approximation that the spectral shape is constant with time and is given by the average rigidity cutoff (12.6 GV) during COSI's orbit (see Figure~\ref{fig:DC3input}). This is a reasonable assumption, considering that $R_{\mathrm{cutoff}}$ only varies from 10.4 to 14.5~GV (see Figure~\ref{fig:Cutoff}). The integrated fluxes of these spectra are used by \textit{cosima} to normalize the light curve.

\subsection{South Atlantic Anomaly (SAA)}

The South Atlantic Anomaly (SAA) is a region where the inner Van Allen radiation belt is closest to Earth's surface. Satellites passing through this area experience high levels of radiation, primarily from protons and electrons. Typically, scientific observations are paused while a satellite traverses the SAA, although some diagnostic data may still be collected. Protons interacting with the satellite can generate both short and long-lived radioisotopes, which decay after the observation resumes, contributing to the overall background. To model the creation and decay of these isotopes, the initial step is to estimate the SAA spectrum along a specific orbit. Since we do not plan to use data taken during SAA passages, we did not simulate the trapped electron component, as these electrons have too low energies ($<$ a few MeV) to induce spallation reactions and only generate prompt emission.

The spectrum of the trapped protons is generated using the model IRENE AP9 v1.57.004\footnote{\url{https://www.vdl.afrl.af.mil/programs/ae9ap9}}~\citep{Ginet2013}. The expected differential flux is computed for every 15s of orbit between March and May 2027 (see Figure~\ref{fig:SAAinputs} for one SAA passage) and then integrated over the energy to get the rate as function of time that MEGAlib will interpret as a light curve (see  Figure~\ref{fig:SAALC}). This light curve will simulate the passage of the satellite through the SAA. For the COSI orbit, the duration of an SAA passage is around 17.5~min. This time is estimated with the condition that the flux of the trapped protons is not equal to zero. One could increase this lower flux boundary to limit the cut in the data but for the purpose of this work we choose to use this strict condition when removing the SAA passages in the simulated data. 

In order to reduce the simulation time, the spectrum is truncated below 4 MeV. This is motivated by the fact that we are only interested in activation induced by the trapped protons during a SAA passage. As shown in Figure~\ref{fig:protoncross}, 
the cross sections for proton-induced reactions with materials commonly found in the spacecraft are negligible below 4 MeV. To test this, a 24h orbit has been simulated for a cut at 2 MeV, showing no significant difference from a cut at 4 MeV.
\begin{figure}[!ht]
    \centering
    \includegraphics[width=\linewidth]{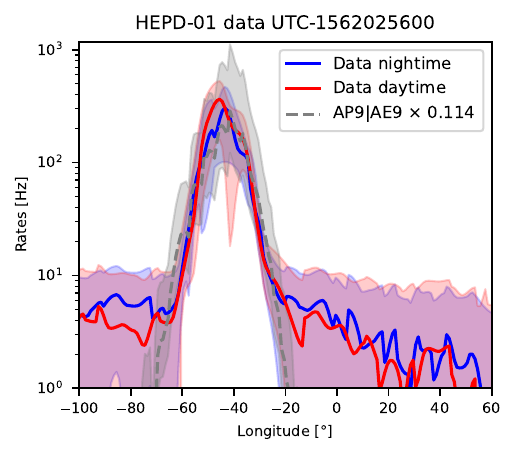}

    \caption{ Comparison between HEPD-01 measurements and AP9/AE9 prediction (grey dashed curve) for a latitude window of [-1°,1°] for one dataset (UTC 1562025600). Shaded areas for both model and data correspond to estimated uncertainties (1~$\sigma$, statistics + systematic). HEPD-01 rates are displayed after subtraction of the estimated cosmic and instrumental baseline. The residual rate outside of the SAA, which is due to the simplicity of the estimation, is nevertheless negligible.  }
    \label{fig:AP9vsdata}
\end{figure}

The instrumental background induced by the trapped protons within the passage of the satellite through the SAA is usually one of the most important background components. The AP9 model shows good agreement with recent measurements by the High-Energy Particle Detector instrument (HEPD-01) on board the China Seismo-Electromagnetic Satellite (CSES-01) in the central part of the SAA~\citep{CSES_SAA}. However, measurements from the BeppoSAX satellite indicate that the AP9 model may strongly overestimate the actual flux values for altitudes below 600 km and inclinations below 5°~\citep{AP9vsdata}. For the work presented in this paper, we studied the comparison between the AP9 model prediction and HEPD-01 data for a latitude window of [-1°,1°]. The altitude of the CSES-01 satellite (500~km) is close to the expected COSI altitude, and likewise for the solar activity (close to minimum).

For this study we selected the central part of HEPD-01 available data, starting from UTC 1562025600 (2 July 2019) to UTC 1640995200 (1 Jan 2022). HEPD-01 data\footnote{HEPD-01 is in sun-synchronous orbit, so that each passage above a given point on the Earth can either occur at daytime (local time 2 PM, descending orbits) or nighttime (local
time 2 AM, ascending orbits).} are averaged within a $\pm$3 month window centered on the given UTC, which gives 6 datasets in total. AP9/AE9 data are extracted within 6 days around the central UTC for each dataset. In this region, HEPD-01 can only provide rate data for a set of trigger masks \citep{CSES_SAA}, so that no identification between the protons and electrons is available. We use data from the trigger mask using an energy cut of 2.3 (29)~MeV for the electron (proton). Given this energy cut, the electron contribution is negligible. A $\chi^2$ fit is employed for each dataset in order to find the best scaling factor $S$ to apply to the AP9/AE9 model to better match the data. Before each fit, we subtracted from HEPD-01 data the baseline rate due to cosmic radiation and instrumental noise. This contribution was estimated by taking the average rate in the longitude region [0°,150°], far away from the SAA. The average scaling factor from each fit gives $S=0.114 \pm 0.047$ (see Figure~\ref{fig:AP9vsdata} for a specific dataset). We thus applied this factor to the predicted flux from AP9 in order to take into account the discrepancy observed by HEPD-01.

 \begin{figure*}[!h]
    \centering
    \includegraphics[trim=8 3 2 2,clip,width=\linewidth]{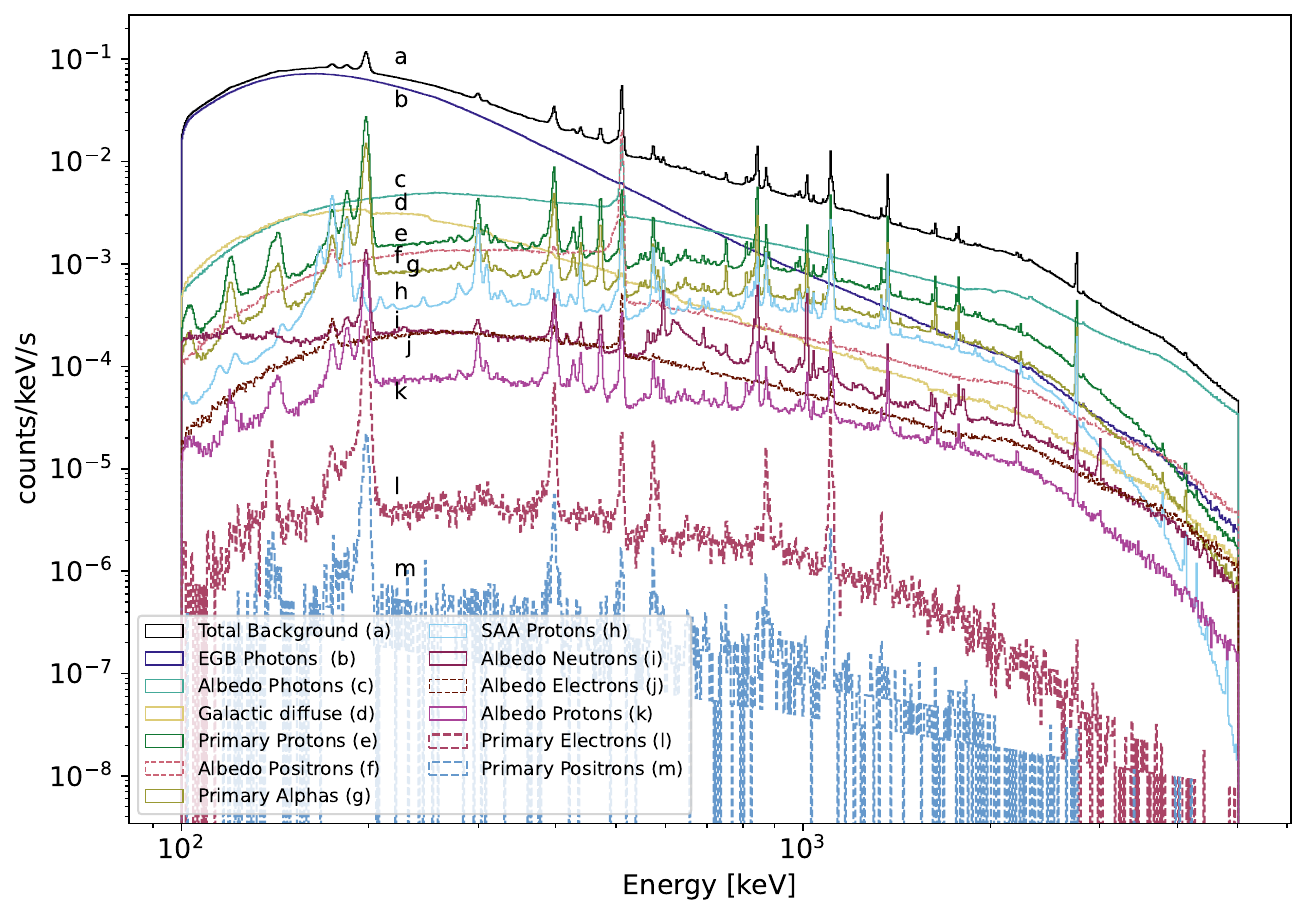}
    \caption{Spectra of the reconstructed Compton events after 3 months of orbit for each background component.}
    \label{fig:DC3_spectra}
\end{figure*}

\begin{figure*}[!ht]
    \centering
    \includegraphics[trim=8 6 8 3,clip,width=0.6\linewidth]{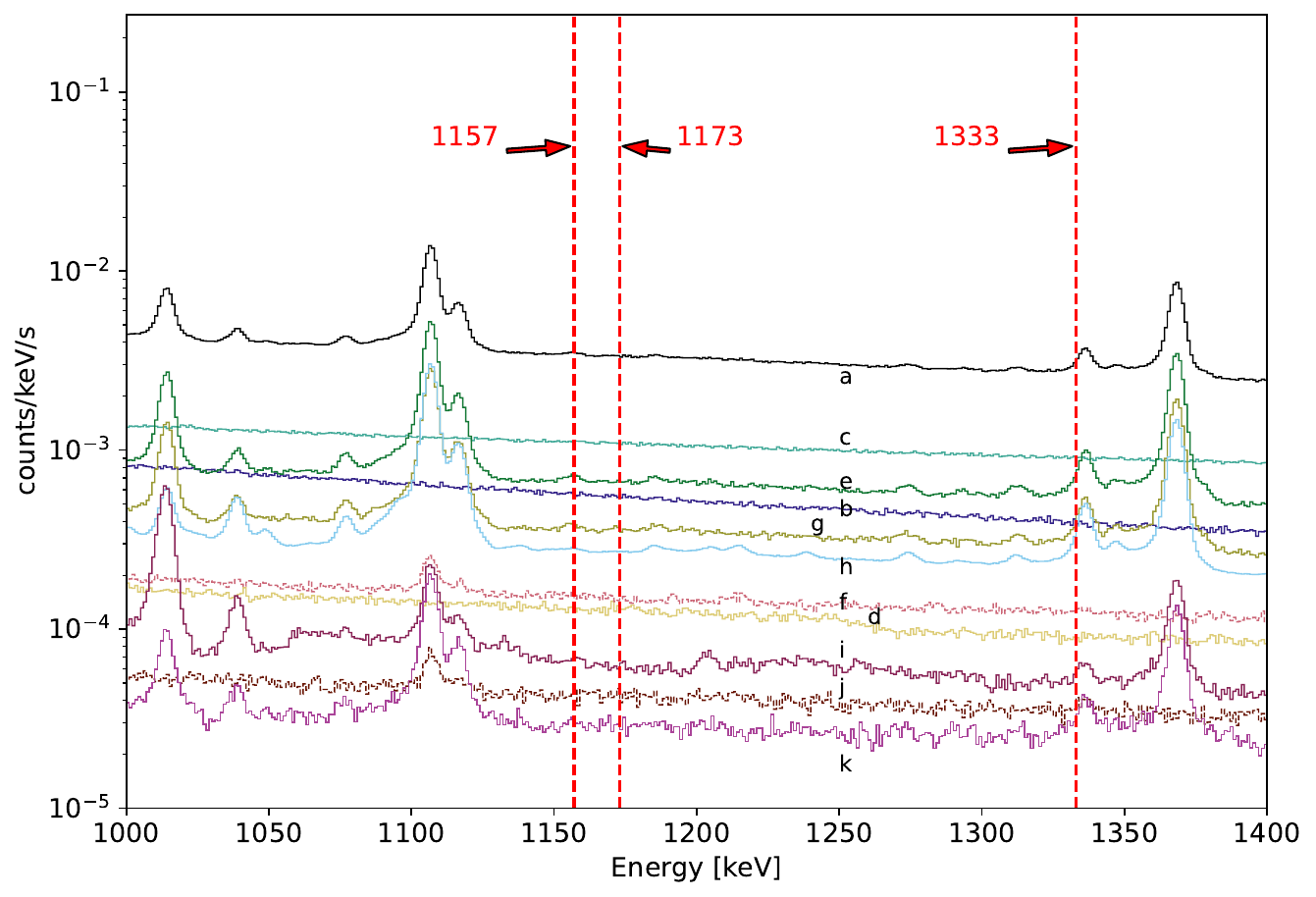}
    \includegraphics[width=0.3\linewidth]{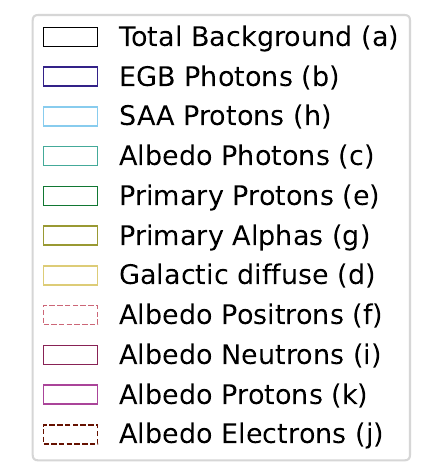}
    \includegraphics[trim=8 6 7.5 5,clip,width=0.48\linewidth]{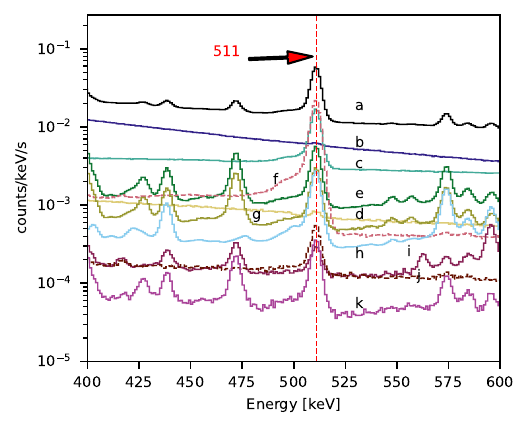}
    \includegraphics[trim=8 6 5 5,clip,width=0.48\linewidth]{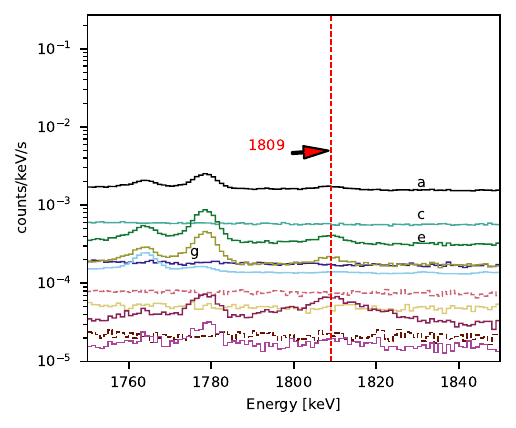}

    \caption{Zoom of Figure~\ref{fig:DC3_spectra} into energy regions of interest (ROI) around gamma-ray lines of COSI science goals, indicated by red-dashed-lines. \textbf{Top:}  ROI around the lines at 1157~keV from $^{44}$Ti and at 1173 and 1333~keV from $^{60}$Fe. 
    \textbf{ Bottom left:} ROI around the 511~keV line from positron annihilation. \textbf{Bottom right:} ROI around the 1809~keV line from $^{26}$Al. Due to their negligible contributions, the primary positrons and electrons components are not represented in these Figures. }
    \label{fig:Spectrazoom2}
\end{figure*}


\section{Results} \label{sec:result}

%

The resulting spectra of the Compton events for each component, as well as the total background, are shown in Figure~\ref{fig:DC3_spectra}. The continuum background is dominated by the extragalactic background photon up to 660~keV, where the albedo photons component takes over. Primary protons and alphas particles contribute to this continuum but their main contribution to the background is from the activation lines they create.

\begin{table}[!ht]

    \centering
    \begin{tabular}{ccc}
    \hline\hline 
         E measured [keV]  & E nominal [keV] & Parent process \\
       \hline
       198.1 & 198.39 & $^{71m}$Ge(IT)$^{71}$Ge \\
       398    &  397.94 & $^{69}$As(EC)$^{69}$Ge \\
       510.8 &  511    & e$^+$ + e$^-$ $\to \gamma + \gamma$\\
       843.8 &   843.76 & $^{27}$Mg($\beta^-$)$^{27}$Al \\
       871.8 & 872.14 & $^{69}$Ge(EC)$^{69}$Ga \\
       1013.9 & 1014.52 & $^{27}$Mg($\beta^-$)$^{27}$Al\\
       1106.9 & 1107.01 & $^{69}$Ge(EC)$^{69}$Ga \\
       1367.1 & 1368.6 & $^{24}$Na($\beta^-$)$^{24}$Mg \\
       2754 & 2754 & $^{24}$Na($\beta^-$)$^{24}$Mg\\
       \hline 
    \end{tabular}
    \caption{Table of the strongest activation lines in the total background spectrum and their parent processes.}
    \label{tab:Activationlines}
\end{table}

The most important lines are listed in Table~\ref{tab:Activationlines}. Overall, the spectrum turns down beyond 2~MeV due to saturation effects in the germanium strips.

Figure~\ref{fig:Spectrazoom2} shows a zoom of Figure~\ref{fig:DC3_spectra} into the energy regions of interest (ROI) around gamma-ray lines of COSI science goals ~\citep{tomsick2023comptonspectrometerimager}. The 511~keV line from e$^+$- e$^-$ annihilation is composed of multiple background components, with the largest contribution from albedo positrons, as depicted in the bottom left panel. The top panel exhibits an activation line at $\sim$1347~keV, close to the 1333~keV line from $^{60}$Fe, another COSI science goal. This background line is due to the reaction $^{69}$Ge(EC)$^{69}$Ga + K + L(1347.1~keV). The bottom right panel shows few counts in the 1809~keV science goal line from $^{26}$Al. Those are blended lines coming from the reactions $^{26}$Na($\beta^-$)$^{26}$Mg(1808.68~keV) and $^{56}$Mn(EC)$^{56}$Fe(1810.772~keV).

\begin{figure*}[!ht]
    \centering
    \includegraphics[width=0.6\linewidth]{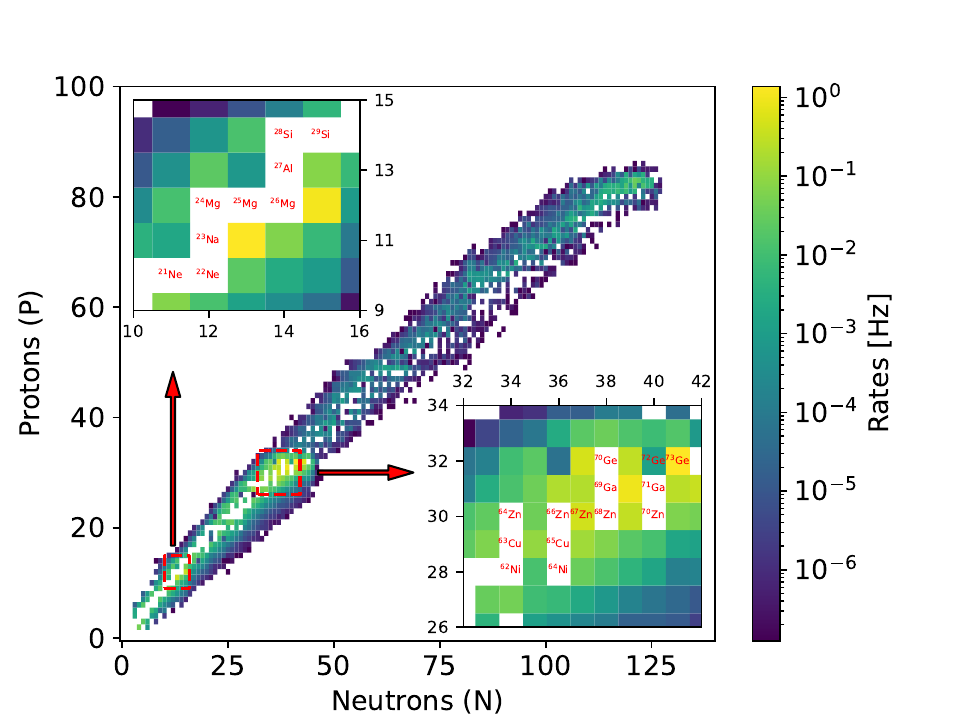}
    \caption{Distribution of isotopes produced by the interaction of the primary protons after 3 months of orbit. The colors represent the Compton events rates associated to these isotopes. Two zooms of this distribution, represented by the red dashed squares, are shown in the upper left and lower right corner. Some stable elements are indicated in red. }
    \label{fig:protonactivation}
\end{figure*}

The similarity of most of the resulting activation lines for different input particles suggests that most isotopes are produced by spallation reactions. A distribution of the isotopes produced by the primary protons is shown in Figure~\ref{fig:protonactivation}. This distribution is derived from all the non-vetoed Compton events that were due to an isotope. We can distinguish three main regions centered around aluminum, germanium, and bismuth, which are primarily found in the spacecraft structure, the detectors themselves, and the anti-coincidence shield, respectively. The 10 most important isotopes, i.e the highest non-vetoed Compton event rates in the [0.2-5]~MeV window, produced by these isotopes, are listed in Table~\ref{tab:PrimaryProtonsactivity}. Some stable elements exhibit non-zero Compton events rates because cosmic rays can excite them to higher energy levels, or because radioactive isotopes can decay into excited states of these stable elements. No contributions of bismuth isotopes are found in Table~\ref{tab:PrimaryProtonsactivity} since the veto of the shield drastically suppresses it. For the primary protons, the BGO shields veto $\sim$92\% of the interactions these protons induce in the Ge detectors. This can reach $\sim$97\% for the primary electrons. This very high veto rate explains the presence of prominent activation lines in the leptonic component, as the prompt continuum contribution, dominated by bremsstrahlung, is drastically suppressed.

\begin{table*}[!ht]

    \centering
    \begin{tabular}{cccc}
    \hline\hline 
         Element & Rate [Hz] & T$_{1/2}$ & $\gamma$/$\beta$ energy [keV]\\
       \hline
       $^{70}$Ga & 1.54 & 21.14 min & \textbf{1651.7}\\       
       $^{75}$Ge/$^{75m}$Ge & 0.87 & 82.78 min/47.7 s & \textbf{1177.2}/139.68 \\                            
       $^{69}$Ge/$^{69m}$Ge & 0.65 & 38.9 h/5.1 us &  574.11, 871.98,\\ & & & 1106.77/397.94\\
       $^{73}$Ga & 0.63 & 4.87 h & \textbf{1234.1},297.32\\
       $^{24}$Na/$^{24m}$Na & 0.59 & 14.96 h/20.18 ms &1368.62,  2754.0/472.2\\
       $^{69}$Zn & 0.50 & 13.756 h & 438.63\\
       $^{72m}$Ga & 0.49 & 39.68 ms & 103.14+16.4\\
       $^{71m}$Ge & 0.48 & 20.22 ms & 174.956+23.438\\
       $^{27}$Mg & 0.45 & 9.458 min & 843.76, 1014.52\\
       $^{68}$Ga & 0.42 & 67.843 min &  \textit{1899.1} , 511, 1077.34 \\

        \hline 
    \end{tabular}
    \caption{The 10 highest rates of Compton events induced by isotopes which are produced by primary protons/alphas and albedo protons/neutrons. The isotope half-lives T$_{1/2}$ and the decay radiation  ($\beta^-$ in bold, $\beta^+$ in italic and $\gamma$ for the rest ), which have a strong impact on the rates are also indicated. 
    For $\beta$ decay, the end-point energy is chosen.}
    \label{tab:PrimaryProtonsactivity}
\end{table*}

\begin{figure*}[!ht]
    \centering
    \includegraphics[width=0.48\linewidth]{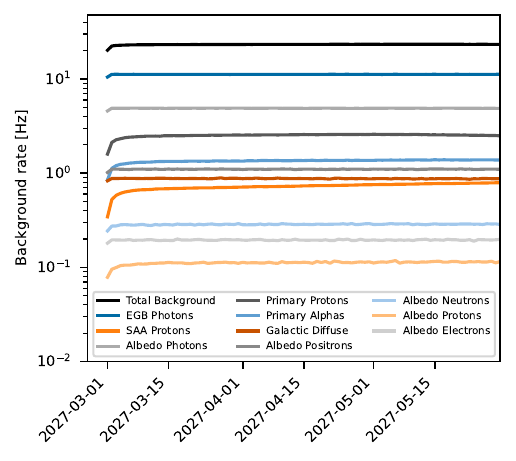}
    \includegraphics[width=0.48\linewidth]{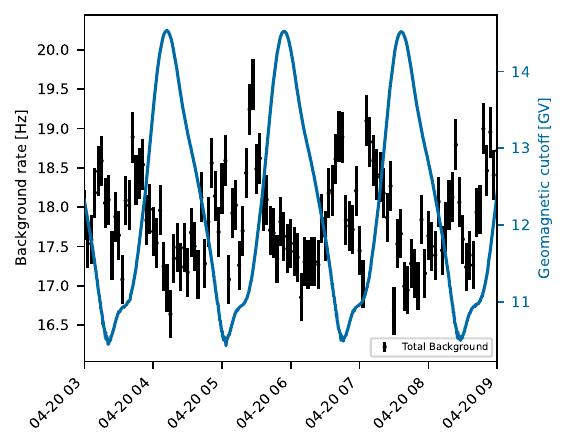}
    \caption{\textbf{Left:} Daily Compton events rates for the different background components, integrated between 0.2 and 5~MeV. Due to their negligible contributions, the Primary positrons and electrons components are not represented in this Figure. \textbf{Right:} The black points show the background rate for every 3 minutes as a function of time for a few hours of orbit. The $R_{\mathrm{cutoff}}$ is shown in blue. The SAA cut is not applied here for better visualization of the rate variation}
    \label{fig:DC3_TotalRate}
\end{figure*}

The integrated rates between 0.2 and 5 MeV as a function of time are shown in the left panel of Figure~\ref{fig:DC3_TotalRate}. On a daily timescale, the Compton event rate remains relatively stable, except during the first few days when it increases before reaching a plateau due to activation build-up. At this scale, variations caused by the geomagnetic cutoff ($R_{\mathrm{cutoff}}$) are difficult to observe. However, on a minute scale, the rate fluctuations become apparent and are inversely correlated with geomagnetic variations, as shown in right panel of Figure~\ref{fig:DC3_TotalRate}. This result in an apparent East/West asymmetry in the simulated background due to the changes in $R_{\mathrm{cutoff}}$. This asymmetry was clearly observed in NuSTAR data~\citep{NuSTAR}.

\subsection{Activation from the SAA}\label{sec:SAAactivation}

\begin{figure*}[!ht]
    \centering
    \includegraphics[width=0.48\linewidth]{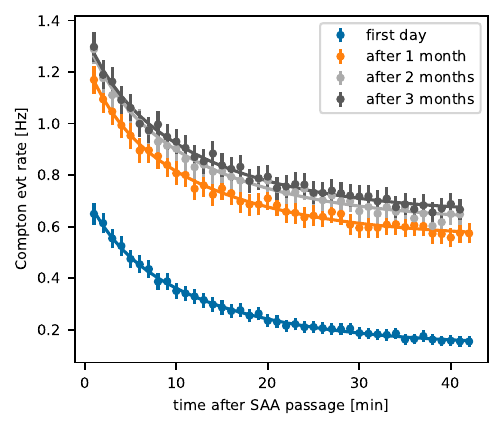}
    \includegraphics[width=0.48\linewidth]{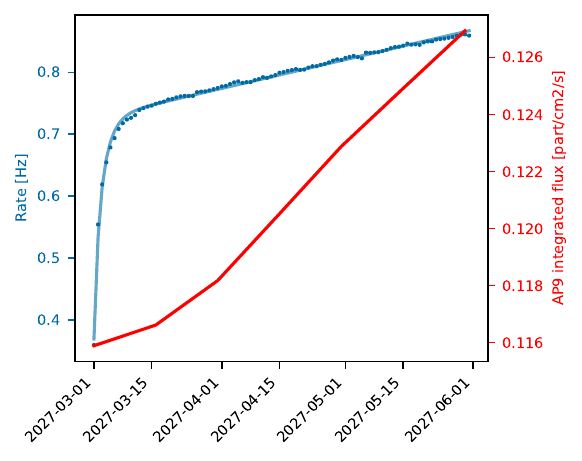}
    \caption{\textbf{Left:} Compton events rate from trapped protons, integrated over (0.2-5)~MeV. Following one SAA passage, after being in orbit for four different durations, as specified in the legend. The solid lines represent the fit of the exponential decay of $^{27}$Mg and $^{28}$Al plus a constant term. \textbf{Right:} Daily rate of Compton events originating from trapped protons during three months of orbit (blue data points). The build-up of the activation is well pronounced during the first days. The blue line represents the fit of the build-up of decays from $^{24}$Na and $^{69}$Ge plus a linear term. The red line shows the integrated AP9 proton flux used as input for the simulation. }
    \label{fig:SAAminscale}
\end{figure*}

 Despite the SAA passage cut, a residual tail can still be observed in the Compton event rate after each SAA passage, as shown in the left panel of Figure~\ref{fig:SAAminscale}. This delayed background is due to the satellite and detector material undergoing nuclear excitation due to collision with the trapped protons, and subsequently photons are produced by the de-excitation of the activated material.

Using a similar approach made with Fermi GBM data~\citep{Biltzinger_2020}, this tail is well fitted with a sum of two exponential decays plus a constant term. On a minute timescale, the delayed background after each SAA passage appears to be mainly driven by the beta decay of $^{27}$Mg and $^{28}$Al. Their corresponding lines at 843.76 and 1014.52~keV for $^{27}$Mg and 1778.987~keV for $^{28}$Al are indeed the most affected when we apply a cool-down period of a few minutes after an SAA passage. 

On the other hand, looking at the Compton event rate for this background component on a daily timescale (see right panel in Figure~\ref{fig:SAAminscale}), we observe a strong build-up during the first days of orbit, followed by a linear increase over time. This linear trend after a few days is mainly driven by the time profile of the AP9 integrated trapped proton flux, as shown in red in the right panel of Figure~\ref{fig:SAAminscale}. Note that this time profile exhibits annual variations, with the flux returning to near the initial value after a full year. 

Similarly to the Compton event rate after one SAA passage, the daily rate is fitted with a sum of two exponentials of the form:
\begin{equation}
    A(1-e^{-t\frac{ln(2)}{T_{1/2}}}),   
\end{equation}
plus a first order polynomial. 
On a daily timescale, the rate increase during the first days appears to be driven by the beta decay of $^{24}$Na, followed by a slower decay from $^{69}$Ge. The isotopes of Na and Mg are coming from the aluminum present in the structure of the spacecraft.


\subsection{Extrapolation to 2 years of orbit}
\begin{figure}[!ht]
     \centering
     \includegraphics[width=\linewidth]{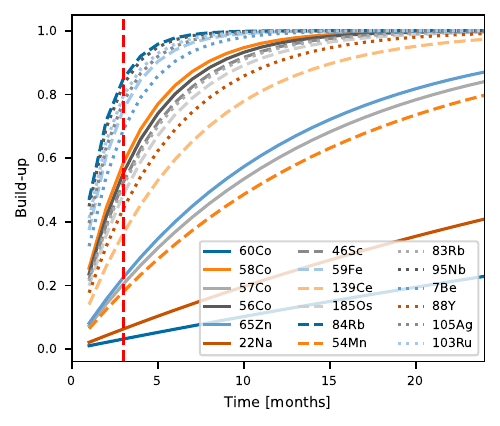}
     \caption{Build-up (B = $1-e^{-t\frac{log(2)}{T_{1/2}}}$) of the long-term activation as a function of time for the 18 most dominant long half-live isotopes. The vertical red-dashed line corresponds to 3 months.}
     \label{fig:longtermbuildup}


\end{figure}

The total computational cost of the simulations reached approximately 3 million CPU~hours (with 1 million just for the SAA~\footnote{The simulation was done using the nominal flux from AP9, so overestimated by a factor $\sim$9.}). Given the magnitude of these computing resources, it was not feasible to extend the simulations beyond the current time frame. Instead, we chose to extrapolate our existing background model to two years in orbit, using the actual outputs from the Monte Carlo simulations. After such a long duration, we expect the contribution from long-lived isotopes to become significant, leading to new emission lines.
Therefore, we identified isotopes with half-lives between 1 month and 10 years which contribute to the background at a measured rate greater than 10$^{-4}$~Hz, finding 18 in total. Figure~\ref{fig:longtermbuildup} shows the expected build-up of their rate as function of time. The extrapolated production rate $R$ after 2 years of orbit inside the volume \textit{i} of the mass-model for the isotope \textit{j} is defined as:
\begin{equation}
 R(2y)^i_j = R(3m)^i_j \times F_j(2y),
 \label{eq:longterm}
\end{equation}

\noindent where $R^i_j(3m)$, is the rate estimated from the 3 month simulation and $F_j(2y)$ is the scaling factor to get the expected rate after 2 years. This factor, defined as the ratio between the production rate after 3 months and 2 years, uses the standard formula for exponential decay, which can be expressed as: 

\begin{equation}
    F_j(2y) = \frac{1-\exp^{-t_{2y}\lambda^j}}{1-\exp^{-t_{3m}\lambda^j}},
\end{equation}

\noindent where $t_{3m}$ and $t_{1y}$ are the times (in seconds) for 3 months and 2 years, and $\lambda^j=\frac{\ln2}{T^j_{1/2}}$ is the decay constant of the isotope j in s$^{-1}$. The corresponding factors for each isotope are listed in Table~\ref{tab:longtermiso}.  

\begin{figure*}[!ht]
    \centering
    \includegraphics[width=0.48\linewidth]{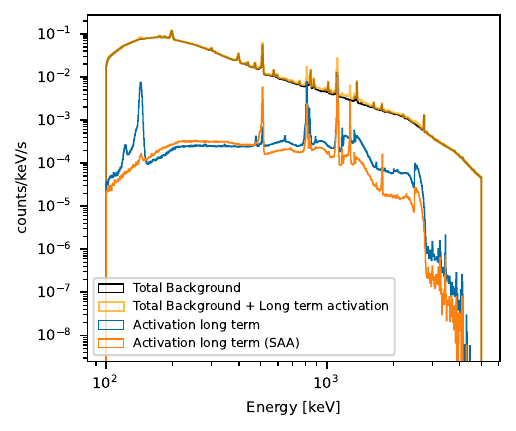}
     \includegraphics[width=0.48\linewidth]{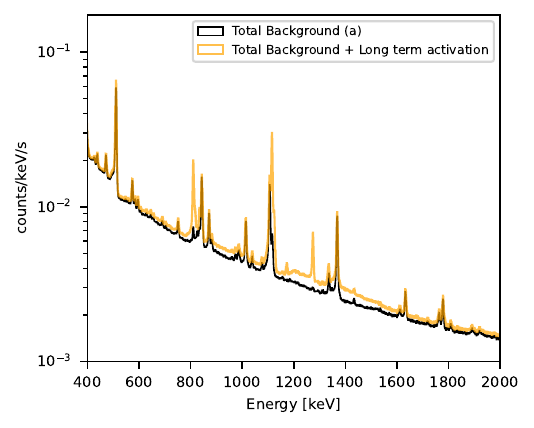}

    \caption{\textbf{Left:} Spectra of the total background model from the 3 month simulations and the expected contribution of the long-lifetime isotopes after 2 years of orbit. \textbf{Right:} Comparison between the background model after three months of orbit and the extrapolated model after two years of orbit between 0.4 and 2~MeV. }
    \label{fig:longtermspectra}
\end{figure*}

Some of the isotopes like $^{22}$Na and $^{60}$Co were already observed and tracked during COMPTEL and INTEGRAL/SPI missions~\citep{SPIlines}. As described in Section~\ref{sec:MEGAlib}, MEGAlib can simulate isotope decay at random positions within the mass model volumes. We thus give as input the list of volumes and isotopes with their extrapolated rates after 2 years of orbit and run the simulation for $1\times10^7$~s in order to get sufficient statistics. The resulting spectra for the Compton events are show in Figure~\ref{fig:longtermspectra}. The overall shape of the background stays relatively constant with some new lines and a build-up of some others like the 511~keV line. The expected rate ratio between 2 years and 3 months of orbit for some energies of interest are provided in Table~\ref{tab:longtermlines}.   

\begin{table}[!ht]

    \centering

    \begin{tabular}{c c c c}
    \hline\hline
       Element & T$_{1/2}$ &F(2y) & $\gamma$ energy [keV]\\ \hline

        $^{60}$Co & 5.27 y & 7.25  & 1173.228, 1332.492\\
        $^{58}$Co & 70.88 d & 1.70 & 511, 810.75\\
        $^{57}$Co & 271.8 d & 4.12 & 122.06, 136.47\\
        $^{56}$Co & 77.23 d & 1.80 & 511, 846.77, 1238.28, \\
                  &           &       &1771.35, 2598.50 \\
        $^{65}$Zn  & 243.93 d & 3.87 & 1115.53\\
        $^{22}$Na & 2.60 y &  6.50 & 511, 1274.53\\
        $^{46}$Sc & 83.80 d & 1.90 & 889.27, 1120.54\\
        $^{59}$Fe & 44.49 d & 1.32 & 1099.24, 1291.59\\
        $^{139}$Ce & 137.64 d & 2.67 & 165.85\\
        $^{185}$Os & 92.95 d & 2.04 & 646.11\\
        $^{84}$Rb & 32.82 d & 1.17 & 551, 881.60\\
        $^{54}$Mn & 312.1 d & 4.43 & 834.84\\
        $^{83}$Rb & 86.2 d & 1.93 & 520.39, 529.59, 552.55\\
        $^{95}$Nb & 34.9 d & 1.20 & 765.80 \\
        $^{7}$Be & 53.3 d & 1.44  & 477.60\\        
        $^{88}$Y & 106.6 d & 2.23 & 898.04, 1836.06\\
        $^{105}$Ag & 41.2 d & 1.28 & 280.54, 344.61,\\ 
                   &         &       &  443.44, 644.63\\
        $^{103}$Ru & 39.2 d & 1.25 & 497.08\\

        \hline
    \end{tabular}
    \caption{Highest contributions from long half-life isotopes in COSI background and their corresponding factor from Equation~\ref{eq:longterm}. The gamma emissions associated with their decays are also indicated.}
    \label{tab:longtermiso}
    \end{table}

\begin{table}[!ht]

    \centering
    \begin{tabular}{ c c c }
    \hline\hline 
        Energy Range [keV] & Science Line [keV] & Rate Ratio  \\
        \hline
        504-516 & 511& 1.101\\
        1155-1159 &  1157 ($^{44}$Ti)& 1.084\\
        1170-1175 &  1173 ($^{60}$Fe)& 1.244 \\
        1331-1335 &  1333 ($^{60}$Fe)& 1.243 \\
        1807-1811 &  1809 ($^{26}$Al)& 1.050 \\
    \hline
    \end{tabular}
    \caption{Ratio of the expected background rate between 2 years and 3 months of orbit for the main COSI science lines.}
\label{tab:longtermlines}
\end{table}

\section{Discussion}\label{sec:Discussion}

\subsection{Comparison with data} \label{sec:datacomp}
A previous study simulated the background for the 2016 balloon flight of the COSI prototype~\citep{gallego2025}. Since the observations were taken at balloon altitudes within Earth's atmosphere, the input models differed; however, because the satellite will use the same type of germanium detectors, we can expect similar activation lines. The simulation from~\cite{gallego2025} successfully reproduced nearly all the activation lines observed in the data, allowing us to distinguish those originating from natural radioactivity (and therefore not included in our model) from those induced by atmospheric background.   

\subsection{Known limitations of our background model}
The current DEE in MEGAlib accounts for detector and readout effects but still at a simplified level. The collaboration is currently developing a more advanced DEE that will be bench-marked with calibration measurements, similar to the COSI-balloon campaign~\citep{Clio_paper,Beechert_2022}. Thus, the rates presented in this paper may differ from the final pre-flight estimates with a fine-tuned DEE.

For this work, the solar activity was assumed to be constant. In reality, the solar activity will vary during the mission and the CR flux is known to be anti-correlated with the solar modulation~\citep{AMS2021}. INTEGRAL/SPI observed an anti-correlation between the sunspot number and rates from activation lines induced by charged particles~\citep{SPIlines}. However, given the high level of modularity of our background model, we will be able to scale each individual component contributions to better represent the change with solar activity.        

We did not simulate the contribution of other ions like oxygen or carbon which are the most important after helium. However, given their low flux in the same order of magnitude as the primary electrons and positrons, we can assume their contributions to the background to be negligible.

Our current model does not take into account activation lines from natural radioactivity. Since some of these lines (mostly from the $^{232}$Th and $^{238}$U series) were observed in the 2016 COSI balloon flight~\citep{gallego2025} and also in space with INTEGRAL/SPI~\citep{SPIlines2,SPIlines}, we can expect to have a significant contribution from it. Background measurements during the calibration campaign will allow us to get an estimation of its contribution prior to launch.



\section{Summary and Conclusion} \label{sec:conclusion}
In this work we simulated the pre-flight background estimates for COSI.
We modeled the background using updated cosmic-ray and albedo spectra, and the Galactic diffuse continuum is modeled using the latest GALPROP models. The simulations account for time-dependent variations due to the geomagnetic cutoff, SAA passages, and they include detailed modeling of delayed activation from short/long-lived isotopes. 

We find that the extragalactic background photon dominates the background at low energies ($<$660 keV), while delayed activation from cosmic-ray primaries (proton/alpha) and albedo photons dominate at higher energies. Even for an equatorial orbit, the delayed activation induced by passage of the satellite through the SAA is non-negligible, especially for the 511~keV line, although taking into account a potential rate overestimation by the AP9 model drastically suppresses its contribution. Time-dependent behavior due to the $R_{\mathrm{cutoff}}$ and SAA-induced activation is quantified. Short-term activation right after SAA passages is dominated by the isotopes produced in the structure of the spacecraft. Our model incorporates a high degree of modularity, enabling us to refine it during the COSI commissioning phase in order to better align with actual measurements.

Given the high computational requirements, simulations were restricted to the first three months of COSI’s orbit. To account for the full mission duration, long-term background levels were extrapolated out to two years—the duration of the prime mission. These results are a critical foundation for optimizing COSI’s observational strategy and refining its background modeling. The background model developed here will be key to enabling detailed studies of COSI's specific science goals. Additionally, this work informed preliminary in-flight calibration plans using activation lines, as described in~\cite{valluvan2025}.   

This work marks a major step forward in understanding the background rates for the COSI satellite mission. These advances are largely made possible by recent improvements to GEANT4 and MEGAlib. As we move closer to launch, our background estimates will become even more accurate through the development of an enhanced DEE and a more detailed mass model, both of which are currently in progress. 

\section*{Acknowledgments}
The Compton Spectrometer and Imager is a NASA Explorer project led by
the University of California, Berkeley with funding from NASA under contract
80GSFC21C0059. Resources supporting this work were provided by the NASA High-End Computing (HEC) Program through the NASA Advanced Supercomputing (NAS) Division at Ames Research Center.

C.M.K.’s research was supported by an appointment to the NASA Postdoctoral Program at NASA Goddard Space Flight Center, administered by Oak Ridge Associated Universities under contract with NASA.

SG and JL acknowledge support by DLR grant 50OO2218. Resources supporting this work were provided by National High Performance Computing (NHR) South-West at Johannes Gutenberg University Mainz.

This work is supported by the Italian Space Agency, partially within contract ASI/INAF No. 2024-11-HH.0.

This work is also supported in part by the Centre National d’Etudes Spatiales (CNES)

HelMod Online calculator (version 4.0.1) website currently supported within the space radiation environment activities of ASIF (ASI - Italian Space Agency -Supported Irradiation Facilities) [2024,07]: www.helmod.org.
The authors want to thank V.Vagelli of the Italian Space Agency for the support on the AMS-02 data interpretation.

This work makes use of data from the CSES mission, a project funded by the China National Space Administration (CNSA), China Earthquake Administration (CEA) in collaboration with the Italian Space Agency (ASI), National Institute for Nuclear Physics (INFN), Institute for Applied Physics (IFAC-CNR), and Institute for Space Astrophysics and Planetology (INAF-IAPS). This work was supported by the Italian Space Agency in the framework of the "Accordo Attuativo 2020-32.HH.0 Limadou Scienza+" (CUP F19C20000110005), the ASI-INFN Agreement No. 2014-037-R.0, addendum 2014-037-R-1-2017, and the ASI-INFN Agreement No. 2021-43-HH.0.

%

\vspace{5mm}






\bibliography{sample631}{}
\bibliographystyle{aasjournal}



\end{document}